\documentstyle[12pt,epsf]{article}
\begin{document}\vspace*{15mm}
\begin{center}{\large \bf Topological chiral symmetry breaking\\[2mm] in SUSY NJL in curved spacetime.}
\end{center}\vspace{8mm}
\begin{center}
L.N. Granda\footnote{e-mail : ngranda@galois.univalle.edu.co }, \\{Departamento de Fisica, Universidad del Valle, A.A. 25360  Cali, Colombia}
\end{center}
\begin{abstract} The effective potential in the model introduced by Buchbinder-Inagaki-Odintsov (BIO) \cite{bio} which represents SUSY NJL model non-minimally coupled with the external gravitational field is found. The topology of the space is considered to be non-trivial. Chiral symmetry breaking under the action of external curvature and non-trivial topology is investigated.\
\end{abstract}\vglue 25mm 
Nambu-Jona-Lasinio (NJL) model represents one of the best solvable examples for the study of dynamical symmetry breaking. SUSY NJL has been introduced in ref. \cite{bul} and it does not contain examples of dynamical symmetry breaking. However, at the external fields such symmetry breaking is possible even in SUSY NJL  as it has been shown recently  in BIO version of the SUSY NJL model non-minimally coupled with external gravity \cite{bio} (for a review of four-fermion models at external fields, see \cite{imo}).\\
In the applications of dynamical symmetry breaking to the early Universe it is important to take into account not only curvature, but also non-trivial topology and temperature effcets. That is why it could be interesting to study dynamical chiral symmetry breaking in SUSY NJL in curved spacetime at non-trivial topology.

That is the purpose of this letter to calculate the effective potential in BIO model in the space ${\cal M}_3$x$S_1$. Two cases are considered in such calculations : when we choose periodic boundary conditions for fermions on the torus $S_1$ and when we choose these conditions to be antiperiodic (that actually corresponds to non-zero temperature case). The numerical study of chiral symmetry breaking under the combined action of curvature and topology or under the action of topology shows the rich phase structure of the BIO model under discussion.

We will start from the action of BIO model \cite{bio} which describes SUSY NJL non-minimally coupled with the external gravity :
\par$$
\begin{array}{rcl}
     S&=&\displaystyle\int d^{4}x\sqrt{-g}\biggl[
         -\phi^{\dag}(\nabla^{\mu}\nabla_{\mu}
         +\rho^{2}+\xi_{1} R)\phi
         -{\phi^{c}}^{\dag}(\nabla^{\mu}\nabla_{\mu}
         +\rho^{2}+\xi_{2} R)\phi^{c}\\
       &&\displaystyle +\bar{\psi}(i\gamma^{\mu}\nabla_{\mu}
         -\rho)\psi-\frac{1}{G}\rho^{2}\biggr].      
\end{array}
\label{s:snjl2}
\eqno{(1)}$$
where $\rho^{2}=\sigma^{2}+\pi^{2}$ is an auxiliary
scalar as in the original NJL model,
$\psi$ is $N_{c}$ component Dirac spinor,
$\xi_{1}$ and $\xi_{2}$ are the non-minimal coupling constants.
The minimal interaction with external gravity 
corresponds to $\xi_{1}=\xi_{2}=0$.
The start point action has the chiral symmetry.
If the auxiliary field $\sigma$ develops the non-vanishing 
vacuum expectation value, $\langle\rho\rangle =m \neq 0$,
the fermion $\psi$ and the scalar $\phi$ acquire
the dynamical mass $m$ and the chiral symmetry could be
broken. From now on we set $\pi=0$, since there is a rotational symmetry in the fields $ \sigma$ and $\pi$, so that we will replace $\rho=\sigma$

In order to calculate the phase structure of the model given by
action (1) we introduce an effective 
potential \cite{bul} :
\par$$
\begin{array}{rcl}
     Z&=&\displaystyle\int {\cal D}\psi{\cal D}\bar{\psi}
         {\cal D}\sigma\ e^{iS}\\
      &=&\displaystyle\int {\cal D}\sigma\frac{\mbox{Det}
         (i\gamma^{\mu}\nabla_{\mu}-\sigma)}
         {\mbox{Det}(\nabla^{\mu}\nabla_{\mu}
         +\sigma^{2}+\xi_{1} R)
         (\nabla^{\mu}\nabla_{\mu}
         +\sigma^{2}+\xi_{2} R)}
         \exp i\int d^{4}x\sqrt{-g}
         \left(-\frac{1}{G}\sigma^{2}\right)\\
      &=&\displaystyle
         \int {\cal D}\sigma\exp i\biggl[\int d^{4}x\sqrt{-g}
         \left(-\frac{1}{G}\sigma^{2}\right)
         -i\mbox{ln Det}(i\gamma^{\mu}\nabla_{\mu}-\sigma)\\
       &&\displaystyle +i\mbox{ln Det}(\nabla^{\mu}\nabla_{\mu}
         +\sigma^{2}+\xi_{1} R)
         +i\mbox{ln Det}(\nabla^{\mu}\nabla_{\mu}
         +\sigma^{2}+\xi_{2} R)
         \biggr].
\end{array}
\eqno{(2)}$$
An internal line of the $\sigma$-propagator gives no contribution
in the leading order of the $1/N_{c}$-expansion.
Assuming that the $\sigma$ is slowely varying field and 
applying the $1/N_{c}$-expansion method the effective potential 
for $\sigma$ is found to be
\par$$
V(\sigma)=\frac{1}{G}\sigma^{2}
+i \mbox{tr}\ln(i\gamma^{\mu}\nabla_{\mu}-\sigma)
-i \mbox{tr}\ln(\nabla^{\mu}\nabla_{\mu}
+\sigma^{2}+\xi_{1} R)
-i \mbox{tr}\ln(\nabla^{\mu}\nabla_{\mu}
+\sigma^{2}+\xi_{2} R)
+{\cal O}\left(\frac{1}{N}\right).
\eqno{(3)}$$
Using the two-point Green functions for scalars and fermions \cite{imo} in Schwinger representation \cite{jul}, the effective potential may be
written as \cite{bio}
\par$$
V(\sigma)=\frac{1}{G}\sigma^{2}
-i \mbox{tr} \int^{\sigma}_{0}ds\ S(x,x;s)
-2i \int^{\sigma}_{0}sds\ [G_{1}(x,x;s)
+G_{2}(x,x;s)]+{\cal O}\left(\frac{1}{N}\right),
\eqno{(4)}$$
where $S(x,x;s)$ and $G_{i}(x,x;s)$ are the spinor and scalar
two-point functions respectively. 
Here it should be noted that the effective potential 
(3) is normalized as $V(0)=0$.
We evaluate the effective potential taking into
account the terms up to linear curvature and
using local momentum representation of propagators on the space ${\cal M}_3$x$S_1$ (see ref. \cite{imo} for a review). The fermion and scalar two point Green functions in this approximation are given in ref. \cite{bio}
\par$$
S(x,x;s)\int\frac{d^{4}p}{(2\pi)^{4}}\left[
\frac{\gamma^{a}p_{a}+s}{p^{2}-s^{2}}
-\frac{1}{12}R\frac{\gamma^{a}p_{a}+s}{(p^{2}-s^{2})^{2}}
+\frac{2}{3}{R}^{\mu\nu}p_{\mu}p_{\nu}
\frac{\gamma^{a}p_{a}+s}{(p^{2}-s^{2})^{3}}\right.
$$
$$
\left.+\frac{1}{4}\gamma^{a}\sigma^{cd}{R}_{cda\mu}p^{\mu}
\frac{1}{(p^{2}-s^{2})^{2}} \right]+{\cal O}(R_{;\mu},R^{2})\, .
\eqno{(5)}$$
\par$$
G_{i}(x,x;s)=\int\frac{d^{4}p}{(2\pi)^{4}}\left[
-\frac{1}{p^{2}-s^{2}}+\left(\frac{1}{3}-\xi_{i} \right)
\frac{R}{(p^{2}-s^{2})^{2}}\right.
$$
$$     \left.-\frac{2}{3}\frac{{R}^{\mu\nu}p_{\mu}p_{\nu}}{(p^{2}-s^{2})^{3}}\right]+{\cal O}(R_{;\mu},R^{2})\, .
\eqno{(6)}$$
Inserting Eqs.(5) and (6)
into Eq.(4) the effective potential becomes \cite{bio}
\par$$
V(\sigma)=\frac{1}{G}\sigma^{2}-4 i \int^{\sigma}_{0}sds
\frac{1}{L}\sum_{n=-\infty}^{\infty}\int \frac{d^{3}p}{(2\pi)^{3}}\left[\frac{1}{p^{2}-s^{2}}
-\frac{1}{12}R\frac{1}{(p^{2}-s^{2})^{2}}\right.
$$
$$
\left.+\frac{2}{3}R^{\mu\nu}p_{\mu}p_{\nu}
\frac{1}{(p^{2}-s^{2})^{3}}\right]+2 i \int^{\sigma}_{0}sds
\frac{1}{L}\sum_{n=-\infty}^{\infty}\int \frac{d^{3}p}{(2\pi)^{3}}\left[
\frac{2}{p^{2}-s^{2}}\right.
$$
$$       
\left.-\left(\frac{2}{3}-\xi_{1}-\xi_{2} \right)
R\frac{1}{(p^{2}-s^{2})^{2}}
+\frac{4}{3}R^{\mu\nu}p_{\mu}p_{\nu}
\frac{1}{(p^{2}-s^{2})^{3}}\right]
\eqno{(7)}$$
For $\xi_{1}+\xi_{2}=1/2$ the chiral symmetry is not broken
down even in curved spacetime. The integration over $s$ is immediate \cite{jul}.Then one should integrate over $p^0$,$p^1$,$p^2$ and sum over the coordinate $p^3$, which is given by $p^3=2\pi n/L$. 

To perform the momentum integration we first make the Wick rotation ($p^0=ip^4$)and put then a cutoff to regularize the resulting expressions. This means the restriction
\par$$
(p^4)^2+(p^1)^2+(p^2)^2\leq\Lambda^2
\eqno{}$$
In the case of purely periodic boundary conditions for both, the fermion and the boson sectors, Eq. (7) simplyfies to
\par$$
V(\sigma)=\frac{1}{G}\sigma^{2}+2 i  \left(\frac{1}{6}-\frac{2}{3}+\xi_{1}+\xi_{2}\right)R
\int^{\sigma}_{0}sds\frac{1}{L}\sum_{n=-\infty}^{\infty}\int \frac{d^{3}p}{(2\pi)^{3}}
\frac{1}{(p^{2}-s^{2})^{2}}+{\cal O}\left(\frac{1}{N}\right).
\eqno{(8)}$$
Note that for $\xi_{1}+\xi_{2}=1/2$ the fermion contribution is cancelled with the boson contribution, conserving in this case the chiral symmetry. After carrying out the integrations over $s$ and the momenta, we  obtain the following expression for the contribution to the effective potential $V(\sigma)$ coming from the purely periodic boundary conditions
\par$$
V(\sigma)=\frac{\sigma^2}{G}-f(\xi_1,\xi_2)\frac{R}{2\pi^2}\frac{1}{L}\sum_{n=-\infty}^{\infty}\left[\frac{2\pi n}{L}\arctan\left(\frac{\lambda L}{2}\right)-\right.
$$
$$
\left. \sqrt{\frac{4\pi^2}{L^2}+\sigma^2}\arctan\left(\frac{\Lambda}{\sqrt{\frac{4\pi^2 n^2}{L^2}+\sigma^2}}\right)\right]
\eqno{(9)}$$
where $f(\xi_1,\xi_2)=\frac{1}{2}-\xi_1-\xi_2$. Performing this sum, using the technique developed in the work \cite{elo}, the effective potential for the model (1) on the background ${\cal M}^3xS^1$ takes the form
\par$$
\frac{V(\sigma)}{\Lambda^4}=\frac{x^2}{g}-f(\xi_1,\xi_2)\left\{\frac{r}{8\pi^2}\left[1-\sqrt{1+x^2}+x^2\mbox{arcsinh}(\frac{1}{x})\right]\right.
$$
$$
\left. +\frac{r}{2\pi^2}\left[\int_0^1 dt\frac{t}{\exp(lt)-1}-\int_x^{\sqrt{1+x^2}}dt \frac{\sqrt{t^2-x^2}}{\exp(lt)-1}\right]\right\}
\eqno{(10)}$$
where $x=\sigma/\Lambda$, $r=R/\Lambda^2$, $l=L\Lambda$, $g=G\Lambda^2$.

Starting from Eq. (10) we can study the dynamical generation of fermion mass. To this end we study numerically the behaviour of the effective potential (10) and look for phase transitions in different situations. As critical parameters we can take $r$, or $l$ which reflects the influence of the topology. \\
In the fixed curvature case we can study the influence of the topology in the phase transition. The results are given in fig.1.
Fixing the parameter $l$ we can study the influence of the curvature in the phase transition. The results are given in fig.2. Note that in both cases we obtained second 
order phase transition.
We now turn to the case of antiperiodic boundary conditions in the fermion sector of Eq. (7). In this case the effective potential takes more complicated form due to the difference in the periodicity of the boundary conditions for fermions and scalars. After integrating over $s$ and the momenta, we obtain
\par$$
V(\sigma)=\frac{\sigma^2}{G}-\frac{1}{L}\sum_{n=-\infty}^{\infty}\left\{\frac{2}{3\pi^2}\left[\sigma^2\Lambda+\frac{(2n+1)^3\pi^3}{L^3}\arctan\left(\frac{\Lambda L}{(2n+1)\pi}\right)\right.\right.
$$
$$
\left.\left. -\left(\frac{(2n+1)^2\pi^2}{L^2}+\sigma^2\right)^{3/2}\arctan\left(\frac{\Lambda}{\sqrt{\frac{(2n+1)^2\pi^2}{L^2}}+\sigma^2}\right)+\frac{\Lambda^3}{2}\log\left(1+\frac{\sigma^2}{\frac{(2n+1)^2\pi^2}{L^2}+\Lambda^2}\right)\right]\right.
$$
$$
\left.+\frac{R}{6(2\pi)^2}\left[\frac{(2n+1)\pi}{L}\arctan\left(\frac{\Lambda L}{(2n+1)\pi}\right)\right.\right.
$$
$$
\left.\left. -\sqrt{\frac{(2n+1)^2\pi^2}{L^2}+\sigma^2}\arctan\left(\frac{\Lambda}{\sqrt{\frac{(2n+1)^2\pi^2}{L^2}}+\sigma^2}\right)\right]\right.
$$
$$
\left.-\frac{R}{6(2\pi)^2}\left[\frac{(2n+1)^2\pi^2 \Lambda}{(2n+1)^2\pi^2+\Lambda^2 L^2}-\frac{((2n+1)^2\pi^2+\sigma^2 L^2)\Lambda}{(2n+1)^2\pi^2+\Lambda^2 L^2+\sigma^2 L^2}\right]\right\}    
$$
$$
+\frac{1}{L}\sum_{n=-\infty}^{\infty}\left\{\frac{2}{3\pi^2}\left[\sigma^2\Lambda+\frac{(2n)^3\pi^3}{L^3}\arctan\left(\frac{\Lambda L}{2n\pi}\right)\right.\right. 
$$
$$
\left.\left. -\left(\frac{(2n)^2\pi^2}{L^2}+\sigma^2\right)^{3/2}\arctan\left(\frac{\Lambda}{\sqrt{\frac{(2n)^2\pi^2}{L^2}}+\sigma^2}\right)+\frac{\Lambda^3}{2}\log\left(1+\frac{\sigma^2}{\frac{(2n)^2\pi^2}{L^2}+\Lambda^2}\right)\right]\right.
$$
$$
\left. -\frac{1}{2\pi^2}(\frac{5}{12}-\xi_1-\xi_2)R\left[\frac{2n\pi}{L}\arctan\left(\frac{\Lambda L}{2n\pi}\right)\right.\right.
$$
$$
\left.\left. -\sqrt{\frac{(2n)^2\pi^2}{L^2}+\sigma^2}\arctan\left(\frac{\Lambda}{\sqrt{\frac{(2n)^2\pi^2}{L^2}}+\sigma^2}\right)\right]\right. 
$$
$$
\left. -\frac{R}{6(2\pi)^2}\left[\frac{(2n)^2 \pi^2 \Lambda}{(2n)^2 \pi^2+\Lambda^2 L^2}-\frac{((2n)^2 \pi^2+\sigma^2 L^2)\Lambda}{(2n)^2 \pi^2+\Lambda^2 L^2+\sigma^2 L^2}\right]\right\}    
\eqno{(11)}$$

To perform the summs in Eq. (11) we may follow the techniques used in \cite{elo}. Nevertheless, because the obtained expression is too large to produce an analytical expresion for the generated fermion mass, we will limit to the case of zero curvature and study the influence of the topology on the symmetry breaking. The numerical results of this study are given in Fig.3. Again the possibility of topology induced phase transition is shown.

In summary, we discussed the phase structure of SUSY NJL model non-minimally interacting with external topologically non-trivial spacetime. The possibility of curvature and (or) topology induced phase transitions are found.
It could be interesting to discuss the applications of our results to static black holes in the same spirit as it has been done for 2D black holes with NJL model in refs. \cite{muo,noj}.
{\bf Aknowledgments} This work has been supported by COLCIENCIAS (Colombia), project No. 1106-05-393-95.

\end{document}